\documentclass[
    twocolumn,
	prd,
	amssymb,
	preprintnumbers,superscriptaddress,
	nofootinbib]{revtex4-1}

\pdfoutput=1
\usepackage{paralist}
\usepackage{graphicx}
\usepackage{enumitem}
\usepackage{latexsym}
\usepackage{amsfonts}
\usepackage{amssymb}
\usepackage{xcolor}
\usepackage[export]{adjustbox}
\usepackage{paralist}
\usepackage{amsmath}
\usepackage[thinlines]{easytable}
\usepackage{slashed}
\usepackage{dcolumn}
\usepackage{verbatim}
\usepackage{float}
\usepackage{multirow}
\usepackage{xspace}
\usepackage[normalem]{ulem}
\usepackage[
pdfauthor={Jeremy Sakstein}]{hyperref}
\usepackage{tabularx}
\usepackage{lettrine}
\usepackage{setspace}

\input Zallman.fd

\LettrineTextFont{\itshape}

\renewcommand{\tilde}{\widetilde} 

\newcommand{\beq}{\begin{equation}}
\newcommand{\eeq}{\end{equation}}
\newcommand{\bea}{\begin{eqnarray}}
\newcommand{\eea}{\end{eqnarray}}

\newcommand{\refjnl}[1]{{\rm#1}}

\def\araa{\refjnl{ARA\&A}}             
\def\apj{\refjnl{ApJ}}                 
\def\apjs{\refjnl{ApJS}}               
\def\aap{\refjnl{A\&A}}                

\def\mnras{\refjnl{MNRAS}}             
\usepackage{pdfbase}[2017/03/16]
\usepackage{xparse,ocgbase}
\usepackage{xcolor,calc}
\usepackage{tikzpagenodes,linegoal}
\usetikzlibrary{calc}
\usepackage{tcolorbox}

\ExplSyntaxOn
\let\tpPdfLink\pbs_pdflink:nn
\let\tpPdfAnnot\pbs_pdfannot:nnnn\let\tpPdfLastAnn\pbs_pdflastann:
\let\tpAppendToFields\pbs_appendtofields:n
\def\tpPdfXform{\pbs_pdfxform:nnnnn{1}{1}{}{}}
\let\tpPdfLastXform\pbs_pdflastxform:
\let\cListSet\clist_set:Nn\let\cListItem\clist_item:Nn
\ExplSyntaxOff

\usepackage{pdfbase}[2017/03/16]
\usepackage{xparse,ocgbase}
\usepackage{xcolor,calc}
\usepackage{tikzpagenodes,linegoal}
\usetikzlibrary{calc}
\usepackage{tcolorbox}

\ExplSyntaxOn
\let\tpPdfLink\pbs_pdflink:nn
\let\tpPdfAnnot\pbs_pdfannot:nnnn\let\tpPdfLastAnn\pbs_pdflastann:
\let\tpAppendToFields\pbs_appendtofields:n
\def\tpPdfXform{\pbs_pdfxform:nnnnn{1}{1}{}{}}
\let\tpPdfLastXform\pbs_pdflastxform:
\let\cListSet\clist_set:Nn\let\cListItem\clist_item:Nn
\ExplSyntaxOff

\makeatletter
\NewDocumentCommand{\tooltip}{%
  ssssO{\ifdefined\@linkcolor\@linkcolor\else blue\fi}mO{yellow!20}mO{0pt,0pt}%
}{{%
  \leavevmode%
  \IfBooleanT{#2}{%
    \ocgbase@new@ocg{tipOCG.\thetcnt}{%
      /Print<</PrintState/OFF>>/Export<</ExportState/OFF>>%
    }{false}%
    \xdef\tpTipOcg{\ocgbase@last@ocg}%
    \ocgbase@add@ocg@to@radiobtn@grp{tool@tips}{\ocgbase@last@ocg}%
  }%
  \tpPdfLink{%
    \IfBooleanTF{#4}{%
      /Subtype/Link/Border[0 0 0]/A <</S/SetOCGState/State [/Toggle \tpTipOcg]>>
    }{%
      /Subtype/Screen%
      /AA<<%
        \IfBooleanTF{#3}{%
          /E<</S/SetOCGState/State [/Toggle \tpTipOcg]>>%
        }{%
          \IfBooleanTF{#2}{%
            /E<</S/SetOCGState/State [/ON \tpTipOcg]>>%
            /X<</S/SetOCGState/State [/OFF \tpTipOcg]>>%
          }{
            \IfBooleanTF{#1}{%
              /E<</S/JavaScript/JS(%
                var fd=this.getField('tip.\thetcnt');%
                if(typeof(click\thetcnt)=='undefined'){%
                  var click\thetcnt=false;%
                  var fdor\thetcnt=fd.rect;var dragging\thetcnt=false;%
                }%
                if(fd.display==display.hidden){%
                  fd.delay=true;fd.display=display.visible;fd.delay=false;%
                }else{%
                  if(!click\thetcnt&&!dragging\thetcnt){fd.display=display.hidden;}%
                  if(!dragging\thetcnt){click\thetcnt=false;}%
                }%
                this.dirty=false;%
              )>>%
            }{%
              /E<</S/JavaScript/JS(%
                var fd=this.getField('tip.\thetcnt');%
                if(typeof(click\thetcnt)=='undefined'){%
                  var click\thetcnt=false;%
                  var fdor\thetcnt=fd.rect;var dragging\thetcnt=false;%
                }%
                if(fd.display==display.hidden){%
                  fd.delay=true;fd.display=display.visible;fd.delay=false;%
                }%
               this.dirty=false;%
              )>>%
              /X<</S/JavaScript/JS(%
                if(!click\thetcnt&&!dragging\thetcnt){fd.display=display.hidden;}%
                if(!dragging\thetcnt){click\thetcnt=false;}%
                this.dirty=false;%
              )>>%
            }%
            /U<</S/JavaScript/JS(click\thetcnt=true;this.dirty=false;)>>%
            /PC<</S/JavaScript/JS (%
              var fd=this.getField('tip.\thetcnt');%
              try{fd.rect=fdor\thetcnt;}catch(e){}%
              fd.display=display.hidden;this.dirty=false;%
            )>>%
            /PO<</S/JavaScript/JS(this.dirty=false;)>>%
          }%
        }%
      >>%
    }%
  }{{\color{#5}#6}}%
  \sbox\tiptext{%
    \IfBooleanT{#2}{%
      \ocgbase@oc@bdc{\tpTipOcg}\ocgbase@open@stack@push{\tpTipOcg}}%
    \tcbox[colframe=black,colback=#7,size=fbox,arc=1ex,sharp corners=southwest]{#8}%
    \IfBooleanT{#2}{\ocgbase@oc@emc\ocgbase@open@stack@pop\tpNull}%
  }%
  \cListSet\tpOffsets{#9}%
  \edef\twd{\the\wd\tiptext}%
  \edef\tht{\the\ht\tiptext}%
  \edef\tdp{\the\dp\tiptext}%
  \tipshift=0pt%
  \IfBooleanTF{#2}{%
    \setlength\whatsleft{\linegoal}%
  }{%
    \measureremainder{\whatsleft}%
  }%
  \ifdim\whatsleft<\dimexpr\twd+\cListItem\tpOffsets{1}\relax%
    \setlength\tipshift{\whatsleft-\twd-\cListItem\tpOffsets{1}}\fi%
  \IfBooleanF{#2}{\tpPdfXform{\tiptext}}%
  \raisebox{\heightof{#6}+\tdp+\cListItem\tpOffsets{2}}[0pt][0pt]{%
    \makebox[0pt][l]{\hspace{\dimexpr\tipshift+\cListItem\tpOffsets{1}\relax}%
    \IfBooleanTF{#2}{\usebox{\tiptext}}{%
      \tpPdfAnnot{\twd}{\tht}{\tdp}{%
        /Subtype/Widget/FT/Btn/T (tip.\thetcnt)%
        /AP<</N \tpPdfLastXform>>%
        /MK<</TP 1/I \tpPdfLastXform/IF<</S/A/FB true/A [0.0 0.0]>>>>%
        /Ff 65536/F 3%
        /AA <<%
          /U <<%
            /S/JavaScript/JS(%
              var fd=event.target;%
              var mX=this.mouseX;var mY=this.mouseY;%
              var drag=function(){%
                var nX=this.mouseX;var nY=this.mouseY;%
                var dX=nX-mX;var dY=nY-mY;%
                var fdr=fd.rect;%
                fdr[0]+=dX;fdr[1]+=dY;fdr[2]+=dX;fdr[3]+=dY;%
                fd.rect=fdr;mX=nX;mY=nY;%
              };%
              if(!dragging\thetcnt){%
                dragging\thetcnt=true;Int=app.setInterval("drag()",1);%
              }%
              else{app.clearInterval(Int);dragging\thetcnt=false;}%
              this.dirty=false;%
            )%
          >>%
        >>%
      }%
      \tpAppendToFields{\tpPdfLastAnn}%
    }%
  }}%
  \stepcounter{tcnt}%
}}
\makeatother
\newsavebox\tiptext\newcounter{tcnt}
\newlength{\whatsleft}\newlength{\tipshift}
\newcommand{\measureremainder}[1]{%
  \begin{tikzpicture}[overlay,remember picture]
    \path let \p0 = (0,0), \p1 = (current page.east) in
      [/utils/exec={\pgfmathsetlength#1{\x1-\x0}\global#1=#1}];
  \end{tikzpicture}%
}

\newcommand{\msun}{{\rm M}_\odot}
 
\newcommand{\mdm}{m_{\rm DM}}

\usepackage{tikz}
\usetikzlibrary{arrows.meta,positioning}

\DeclareRobustCommand{\okina}{%
  \raisebox{\dimexpr\fontcharht\font`A-\height}{%
    \scalebox{0.8}{`}%
  }%
}

\newcommand{\vtwo}[1]{\textcolor[rgb]{0.,0,0.}{#1}}

\allowdisplaybreaks

\begin{document}

\title{Dark matter silences Cepheids in the Galactic Center}

\author{Djuna Croon} \email{djuna.l.croon@durham.ac.uk}
\affiliation{Institute for Particle Physics Phenomenology, Department of Physics, Durham University, Durham DH1 3LE, U.K.}

\author{Tim Linden}
\thanks{{\scriptsize Email}:~\href{mailto:linden@fysik.su.se}{linden@fysik.su.se};  \href{http://orcid.org/0000-0001-9888-0971}{0000-0001-9888-0971}}
\affiliation{Stockholm University and The Oskar Klein Centre for Cosmoparticle Physics, Alba Nova, 10691 Stockholm, Sweden}
\affiliation{Erlangen Centre for Astroparticle Physics (ECAP), Friedrich-Alexander-Universität \\ Erlangen-Nürnberg, Nikolaus-Fiebiger-Str.~2,
91058 Erlangen, Germany}

\author{Jeremy Sakstein} \email{sakstein@hawaii.edu}
\affiliation{Department of Physics \& Astronomy, University of Hawai\okina i, Watanabe Hall, 2505 Correa Road, Honolulu, HI, 96822, USA}

\date{\today}

\begin{abstract}
Upcoming near–infrared facilities (e.g.\ JWST/NIRCam, ELT/MICADO) will dramatically increase the detectability of galactic center Cepheids despite extreme extinction at optical wavelengths.~In this work, we study the impact of dark matter (DM) annihilation on Cepheid stars in the inner parsec of the Milky Way.~We show that at \vtwo{captured} densities $\rho\sim10^5\,$GeV\,cm$^{-3}$, blue loop evolution can be suppressed, preventing the formation of low-mass ($3-6~\msun$) short–period ($1$--$6$ days) Cepheids.~\vtwo{For even slightly higher DM densities, Cepheids are suppressed across their entire mass range.} A dearth of such variables could provide indirect evidence for DM heating.~Notably, this effect occurs at lower DM densities than required to impact main–sequence stars.~Future surveys will thus offer a novel, complementary probe of DM properties in galactic nuclei.
\end{abstract}

\preprint{IPPP/25/61}

\maketitle

\section{Introduction}

Discovering the particle nature of dark matter (DM) remains one of the paramount goals of astrophysics and cosmology.~A wide range of DM candidates have been put forward, presenting a challenge for identifying the observational signatures of each.~A particularly well-motivated and generic scenario is thermal freezeout in the early universe, which requires that DM particles scatter and annihilate with Standard Model (SM) particles.~Such interactions can be probed through their impact on stellar evolution~\cite{Batell:2009zp,Pospelov:2007mp,Rothstein:2009pm,Pospelov:2008jd,Chen:2009ab,Schuster:2009au,Schuster:2009fc,Bell_2011,Feng:2015hja,Kouvaris:2010,Feng:2016ijc,Allahverdi:2016fvl,Leane:2017vag,Arina:2017sng,Albert:2018jwh,Albert:2018vcq,Nisa:2019mpb,Niblaeus:2019gjk,Cuoco:2019mlb,Serini:2020yhb,Acevedo:2020gro,Mazziotta:2020foa,Bell:2021pyy,Bose:2021cou,Smirnov:2022zip,Croon:2023trk,John:2024thz,Croon:2024waz}.~In this work we study the impact of DM on Cepheid variable stars, considering WIMP models where captured particles settle in the stellar core and annihilate, injecting additional energy into the star.

\begin{figure}[t]
    \centering
    \includegraphics[width=0.48\textwidth]{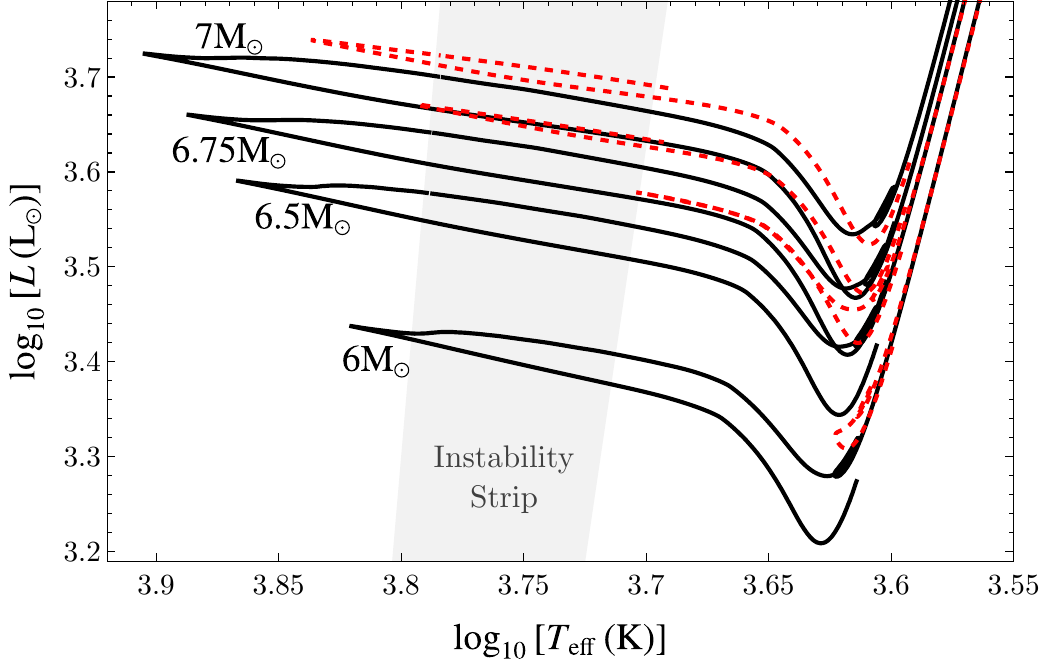}
    \caption{Vanishing blue loops:
    the evolution of $6$--$7~\msun$ stars in the SM (black, solid) and with DM annihilation (red, dashed) with $m_{\rm DM}=1$ GeV, $\vtwo{f_{\rm cap}}\rho_{\rm DM}=10^5$ GeV/cm$^3$, and $v_{\rm DM}= 50$ km/s .~The gray region shows the instability strip.
    }
    \label{fig:hold_my_beer}
\end{figure}

Cepheids are a class of stars that undergo regular periodic pulsations.~They have masses spanning 3--12$~\msun$ with periods of 1-50 days.~They have previously been found to be excellent probes of new physics \cite{Jain:2012tn,Friedland:2012hj,Sakstein:2016lyj,Anderson:2024sfq}.~As explained in more detail below, during their evolution these stars execute \textit{blue loops} in the Hertzsprung–Russell (HR) diagram, shown in Figure~\ref{fig:hold_my_beer}, where they move blueward (to higher effective temperature) at nearly constant luminosity before turning redward again.~While on this sojourn they cross the \textit{instability strip}, a region where opacity variations drive large-amplitude pulsations through the \textit{$\kappa$-mechanism}.~

In this work, we simulate the evolution of Cepheids under the influence of DM annihilation.~We find that the blue loops of low-mass ($3$--$6~\msun$) Cepheids, which have shorter periods ($1$-$6$ days) are suppressed if $\vtwo{f_{\rm cap}}\rho_{\rm DM}\sim10^5$ GeV/cm$^3$ \vtwo{(where $f_{\rm cap}$ is the capture fraction),} and are \vtwo{absent if the captured DM density is slightly larger.}~An example is shown in Figure~\ref{fig:hold_my_beer}.~The extent of suppression depends upon the DM mass and interaction cross-section.~This implies that  Cepheids might not form in the inner parsecs of the Milky Way, where the DM density can be extremely high~\cite{Diemand:2008in, Navarro:2008kc, Bertone:2024wbn}.~

While no Cepheids are currently observed near the galactic center (GC), a result that is consistent with our predictions, this may also be due to observational limitations such as crowding.~Ongoing and planned surveys, e.g., JWST/NIRCam and ELT/MICADO, are expected to observe these objects --- if they exist.~The continued absence of short-period Cepheids would provide a distinctive signature of DM.

This work is organized as follows.~In Section \ref{sec:Cepheid_intro} we review the physics of Cepheid stars.~We discuss our models of DM annihilation in Section~\ref{sec:DM_injection}.~We describe our stellar simulations in Section~\ref{sec:stellar_response}.~The results of our simulations are presented and explained in Section~\ref{sec:blue_loops_DM}.~We discuss prospects for testing DM using the absence of short-period Cepheids in the galactic center in Section~\ref{sec:discussion}.

\section{Cepheid Stars}
\vspace{-0.25cm}

\label{sec:Cepheid_intro}
\begin{figure}
    \centering
    \includegraphics[width=0.45\textwidth]{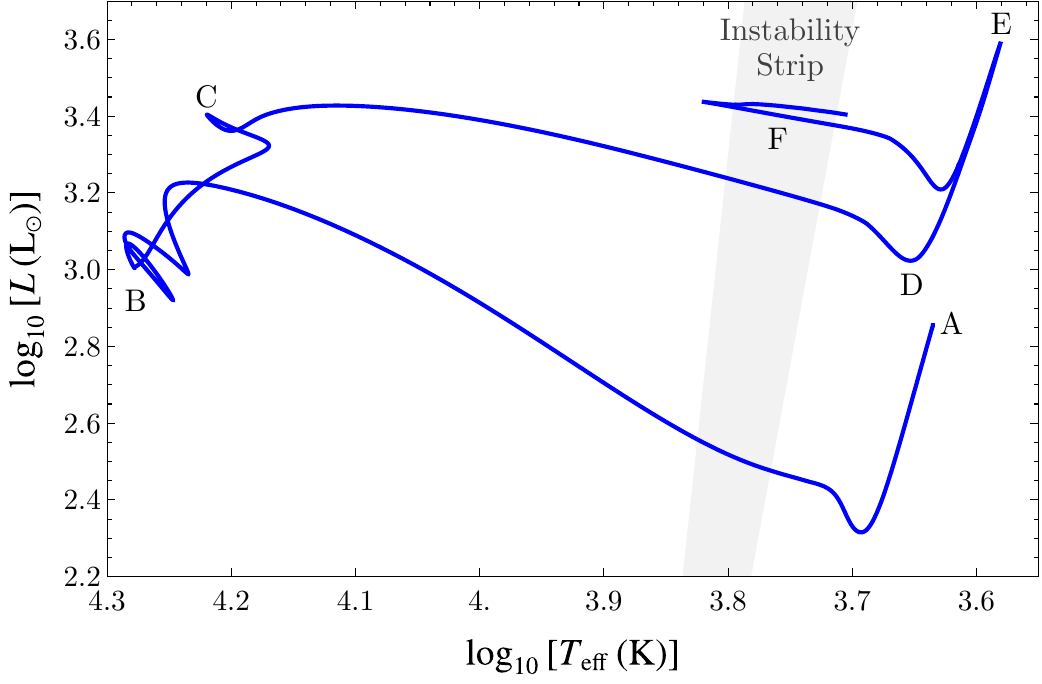}
    \caption{Hertzsprung-Russell Diagram of the evolution of a $6~\msun$ star with metallicity $Z=0.02$ and initial helium abundance $Y=0.29$, consistent with galactic center stars.~The gray region is the Cepheid instability strip.}
    \label{fig:cepheid_evolution}
\end{figure}

The evolution of Cepheid stars is shown in Figure~\ref{fig:cepheid_evolution}, using an example mass of $6~\msun$.~The star begins as a collapsing dust cloud at point A.~Its core temperature and density increase until hydrogen is ignited at point B, heralding the Zero-Age Main Sequence (ZAMS).~The core hydrogen is depleted at the terminal-age main-sequence (point C), leaving behind an inert helium core surrounded by a hydrogen-burning shell, which provides the star's luminosity.~The shell drives an expansion of the envelope, causing the star to cool rapidly and move across the Hertzprung gap (transition from C to D).~The cooling leads to an increase in the opacity, inhibiting radiative transport and making the envelope increasingly convective.~At point D the convective envelope reaches its maximum depth, mixing processed material to the surface in the \textit{first dredge-up}.~The star then ascends the Hayashi line.~The shell deposits helium ash on the core, increasing its temperature until it is sufficient for helium burning, which begins at point E.~The star then descends the Hayashi track and executes a \textit{blue loop} marked F.~The blue loop is a result of excess helium above the hydrogen burning shell \cite{2015MNRAS.447.2951W}.~This excess causes the star to be redder than expected.~As the H-burning shell moves outwards, the excess is smaller, causing the loop.

The gray shaded region is the \textit{Cepheid instability strip}.~Stars evolving through this region pulsate with large amplitudes, producing periodic variations in their bolometric luminosity (not shown on the HR diagram).~The pulsation period follows the period–mean density relation, $P\sim \sqrt{R^3/GM}$, and the amplitude increases from $\sim$10\% in short-period Cepheids to $\sim$60\% in long-period objects.~The pulsations are driven by the \textit{$\kappa$-mechanism}.~A star becomes unstable to pulsations whenever its opacity in a radial layer increases upon compression.~This produces a cycle that operates as follows:~(1) the increasing opacity in the compressed layer traps radiation more efficiently in the layers below, (2) heat builds up in the lower layers which increases the pressure and pushes the initial layer outwards, (3) this causes the density and opacity of the initial layer to decrease, releasing the trapped radiation and restarting the cycle.

In most stars, the opacity decreases upon compression and so the $\kappa$-mechanism is not active.~For example, Kramer's opacity law, $\kappa\propto \rho/T^{3.5}$, implies that opacity is more sensitive to variations in temperature than density so most compressions act to reduce $\kappa$ because they increase the temperature.~An exception is partial-ionization zones where the energy from the compression is used to further ionize the layer and the temperature is approximately unchanged.~In Cepheids, a layer of partially-ionized helium in the envelope drives the pulsations via the $\kappa$-mechanism.~The instability strip is the region of the HR diagram where the He-ionization zone can drive pulsations efficiently.~In stars cooler than the red (right) edge, the He+ zone is deep enough that convection can efficiently transport energy across it, and the $\kappa$-mechanism in ineffective.~In stars hotter than the blue (left) edge the He+ zone is close to the surface and does not contain enough mass to drive pulsations.

\vtwo{Several standard stellar physics uncertainties can affect the existence and extent of blue loops; we will briefly review the most important effects. Firstly, increasing the metallicity raises the opacity and reduces loop extent. The range $Z=0.02\text{--}0.04$, relevant for the galactic center, changes the loop temperature by $\delta\log_{10}(T_{\rm eff}/{\rm K})\sim0.1$ \cite{2004A&A...418..225X,2015MNRAS.447.2951W}. Helium enrichment ($Y\simeq 0.28\text{--}0.32$ for $\Delta Y/\Delta Z=2\text{--}2.5$ \cite{Aghanim:2018eyx,2007MNRAS.382.1516C}) similarly increases the opacity and suppresses loops, but gives a smaller shift in temperature: $\delta\log_{10}(T_{\rm eff}/{\rm K})\sim0.05$ \cite{2015MNRAS.447.2951W}. Uncertainties in the opacity itself are only $\sim5\%$ \cite{Saltas:2022aua} and induce shifts an order of magnitude smaller than those from composition. Models with too little core overshooting yield a He core too small to drive a loop, while too much erases the H/He discontinuity required for loop formation \cite{2004A&A...418..213X,2014arXiv1410.1652H,Wagle:2019vok,Sakstein:2019qgn,Anderson:2024sfq}. The mixing length also plays a role: larger $\alpha_{\rm MLT}$ smooths the envelope gradient and suppresses loops, whereas smaller values preserve the gradient and enhance them \cite{2014MNRAS.445.4287T}. Semiconvection helps establish the required H/He composition gradient, though the loop extent depends only weakly on its efficiency \cite{1994A&A...282..843M,1995MNRAS.275..983E}. We expand on our modeling choices in Section.~\ref{sec:stellar_response}.}

\section{Dark matter energy injection}
\label{sec:DM_injection}
\vspace{-0.25cm}
\noindent We model energy injection from DM annihilation using the standard capture–annihilation framework.~Evaporation is neglected, which is valid for the DM masses of interest ($m_{\rm DM}\gtrsim $ GeV) and the objects we consider \cite{Garani:2021feo}.~The essential ingredients are capture, annihilation, thermalization, and diffusion, which we treat as follows.

\subsection{Capture and annihilation equilibrium}

\begin{figure}
    \centering
    \includegraphics[width=0.85\linewidth]{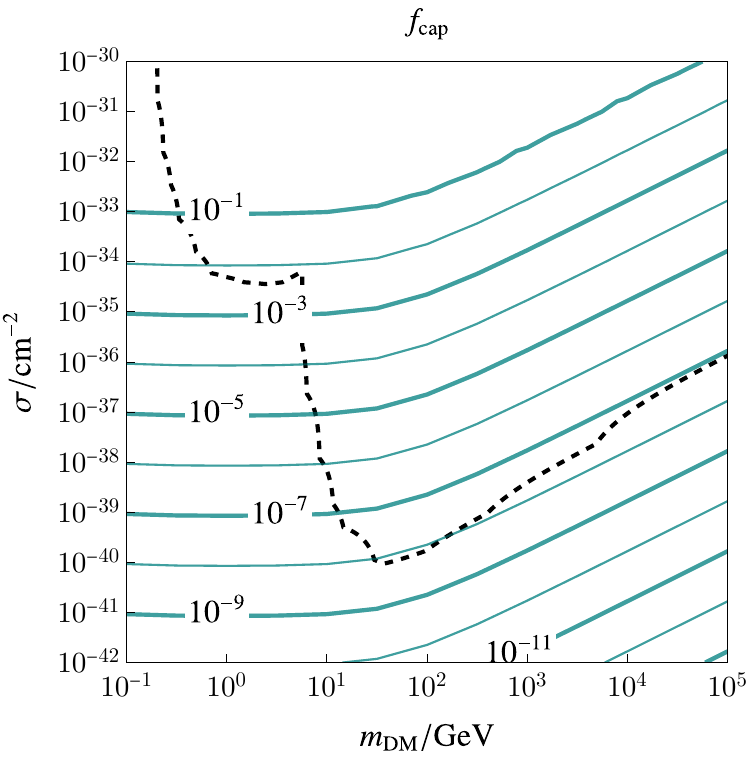}
    \caption{Capture fraction for a $5.5~\msun$ Cepheid of radius $50 \,R_\odot$, assuming spin-dependent interactions and $70.6\%$ hydrogen by mass, computed using Asteria \cite{Leane:2023woh}.~Dashed lines indicate direct detection constraints from CRESST \cite{CRESST:2022dtl} and LUX \cite{LUX:2017ree}.} 
    \label{fig:fcap}
\end{figure}

We first assume that capture and annihilation reach equilibrium on timescales that are short compared to stellar evolution.~The equilibration time is
\begin{equation}
    \tau_{\rm eq} \sim \left( \Gamma_{\rm ann} \Gamma_{\rm cap} \right)^{-1/2},
\end{equation}
where the capture rate is estimated as $\Gamma_{\rm cap} = \Phi \pi R^2$, with DM flux density \cite{Leane:2022hkk}
\begin{equation}
\label{eq:DMflux}
    \Phi = v_{\rm DM} \sqrt{\frac{8}{3\pi}} \left[ 1 + \frac{3}{2}\left(\frac{v_{\rm esc}}{v_{\rm DM}}\right)^2 \right] \frac{\rho_{\rm DM} f_{\rm cap}}{m_{\rm DM}} .
\end{equation}
Here $f_{\rm cap}$ is the fraction of DM captured as it streams through the star and $\rho_{\rm DM}$ is the ambient DM density.~The annihilation rate per effective volume is $\Gamma_{\rm ann} =  \langle \sigma_{\rm ann} v \rangle / V_{\rm eff}$, with $V_{\rm eff} = 4\pi R^3/3$.~For a representative $5\,M_\odot$ star of radius $R_\odot$, the equilibrium timescale is:
\begin{equation}
\begin{split}
    \tau_{\rm eq} \sim& \, 20 \, {\rm yr}\,
    \sqrt{\frac{1}{f_{\rm cap}} 
    \frac{ m_{\rm DM}\, 10^{-30} \,\text{cm}^3\text{/s} }{ \langle \sigma_{\rm ann} v\rangle \,\text{GeV} }}
    \sqrt{\frac{\rho_{\rm DM}}{10^4 \,\rm GeV \, cm^{-3}}} ,
\end{split}
\label{eq:taueq}
\end{equation}
which is negligible compared to stellar lifetimes.~In what follows, we set $f_{\rm cap}=1$ (maximal capture), with results for smaller fractions obtained by rescaling $\rho_{\rm DM}$.~Figure~\ref{fig:fcap} shows the scaling of $f_{\rm cap}$ with DM mass and scattering cross section.

\subsection{Thermalization and transport}

We assume that captured DM thermalizes with the stellar medium.~The thermalization time is the number of scatterings needed for the DM to enter thermal equilibrium (set by the DM–SM mass ratio), multiplied by the mean free path, $\lambda = (\sigma_0 n_{\rm SM})^{-1}$, divided by the velocity dispersion (taken as the escape velocity) \cite{Iocco:2008rb}:
\begin{equation}
\begin{split}
    \tau_{\rm th}(r) =& \frac{m_{\rm DM}}{\rho_{\rm SM}(r) \sigma_0} \sqrt{\frac{r}{2 G M}} \\
        =& 7.2 \, {\rm yr} \left(\frac{m_{\rm DM}}{1 \rm GeV} \right)
        \left( \frac{\sigma_0}{10^{-35} \rm cm^{2}}\right)^{-1} \\&\times
        \left( \frac{R}{50 R_\odot}\right)^{7/2}
        \left(\frac{M_\star}{5 M_\odot} \right)^{-3/2}.
    \end{split}
    \label{eq:tauth}
\end{equation}
For Cepheid parameters, $\tau_{\rm th}$ is orders of magnitude shorter than evolutionary timescales, making this a safe assumption.\footnote{This timescale is longer than the pulsation period, but this does not impact our conclusions since our interest lies in whether stars cross into the instability strip, not with the effects of DM on the pulsations themselves.}~We also assume diffusion is rapid compared to stellar evolution.~Estimating the diffusive timescale as:
\begin{equation}
    t_{\rm diff} \sim \frac{(\Delta x)^2}{\lambda v_T}
\end{equation}
with $\Delta x\sim 50\,R_\odot$, $n_{\rm SM}\sim 10^{19}\,\rm cm^{-3}$, $\sigma_0\sim 10^{-35}\,\rm cm^2$, and $v_T \sim 12\,\rm km\,s^{-1}$, we obtain $t_{\rm diff}\sim 17$ min \cite{Leane:2022hkk}, which is again negligible compared to stellar times.~The advective timescale across the envelope is of order one year ($\Delta x/v_{\rm bf}$ where $v_{\rm bf} \sim 10^5\,\rm cm\,s^{-1}$ is the bulk flow speed), still small compared to Cepheid evolutionary phases.~This ensures that the heat generated by DM annihilation is transported efficiently.

\subsection{Spatial distribution and energy injection}

The DM spatial distribution depends on the Knudsen number $K=\lambda/r_\chi$, with $r_\chi = \sqrt{3k_B T_c/(2\pi G \rho_c m_{\rm DM})}$ (see e.g.~\cite{Banks:2021sba}).~In the small-$K$ limit (frequent scatterings), the Local Thermal Equilibrium (LTE) profile is given by a Maxwell–Boltzmann distribution \cite{Gould:1989hm}:
\begin{equation}\label{eq:MBdist}
    \left(\frac{n_{\rm DM} (r) }{n_{\rm DM}(0)} \right)_{\rm LTE}=
    \left(\frac{T(r)}{T(0)} \right)^{3/2} 
    e^{- \int_0^r d\tilde{r} 
    \frac{\alpha(\tilde{r})dT/d\tilde(r) + m_{\rm DM} g(\tilde{r}) }{T(\tilde{r})}
    } ,
\end{equation}
where $g(r)$ is the gravitational acceleration and $\alpha(r)$ is related to the diffusion coefficient \cite{Leane:2022hkk}.~In the approximation of $\alpha$ constant, one obtains the simpler form \cite{Leane:2022hkk}
\begin{equation}\label{eq:MBdist2}
    \left(\frac{n_{\rm DM} (r) }{n_{\rm DM}(0)} \right)_{\rm LTE}= 
    \left(\frac{T(r)}{T(0)} \right)^{-1 + \frac{1}{2} \left(1+\frac{m_{\rm DM}}{m_{\rm SM}}\right)^{-3/2}} 
    e^{- \int_0^r d\tilde{r} 
    \frac{ m_{\rm DM} g(\tilde{r}) }{T(\tilde{r})}
    } .
\end{equation}

In the large-$K$ limit (rare scatterings), the distribution approaches the isothermal form \cite{Spergel:1984re}
\begin{equation}
    \left(\frac{n_{\rm DM} (r) }{n_{\rm DM}(0)} \right)_{\rm iso}
    = \exp\!\left[-r^2/r_\chi^2\right].
\end{equation}
We interpolate between the two limits as \cite{Bottino:2002pd,Scott:2008ns,Gould:1989hm}
\begin{equation}
    n_{\rm DM} (r)  = f(K) n_{\rm DM}^{\rm LTE} (r) + (1-f(K)) n_{\rm DM}^{\rm iso} (r) ,
    \label{eq:ndm_interpol}
\end{equation}
with $f(K) = 1 - (1+(K/K_0)^2)^{-1}$ and $K_0 = 0.4$.~The DM profile is thus determined by the DM mass, cross section, and ambient density, together with the stellar core temperature and density.~For the cross sections of interest here ($\sigma_0 \lesssim 10^{-30}\,\rm cm^{2}$), Cepheids fall in the isothermal regime.~We fix to be in this regime in what follows.

The energy injection per unit mass is
\begin{equation}
\label{eq:energyinjection}
    \epsilon_{\rm DM}(r) = f_\nu \frac{\langle \sigma v \rangle n_{\rm DM}^2(r) m_{\rm DM}}{\rho(r)} ,
\end{equation}
where $\rho(r)$ is the stellar density and $f_\nu$ accounts for the fraction of annihilation products that deposit their energy locally.~Here, we take $f_\nu=1$ to find the maximal effect.~In annihilation equilibrium the DM luminosity is:
\begin{equation}
    L_{\rm DM} = 4 \pi \int_0^{R} \rho(r)\,\epsilon_{\rm DM}(r)\, r^2 dr = m_{\rm DM} \Gamma_{\rm cap},
    \label{eq:lumanneq}
\end{equation}
demonstrating that the net luminosity becomes independent of $\langle \sigma v \rangle$ \cite{Lopes:2021jcy}.~We normalize the profile~\eqref{eq:energyinjection} accordingly using Eq.~\eqref{eq:lumanneq}.

\section{Stellar Structure Code}
\label{sec:stellar_response}
We modified the stellar structure code MESA (version 23.05.1)~\cite{Paxton:2010ji,Paxton:2013pj,Paxton:2015jva,Paxton:2017eie,2019ApJS..243...10P,2023ApJS..265...15J} to include dark matter energy injection from the profile in Equation~\eqref{eq:ndm_interpol}.~This was accomplished by using the {\tt other\_energy\_implict} hook.~The full details of our implementation and our MESA inlists can be found in a reproduction package that accompanies this paper:~\href{https://zenodo.org/records/17254126}{https://zenodo.org/records/17254126} \cite{sakstein_2025_17254126}.~We briefly summarize the salient features of our stellar modeling choices here.

\indent $\bullet{}$~\textbf{Semi-Convection:}~We use the Langer scheme \cite{1985A&A...145..179L} with efficiency $\alpha_{\rm sc }=0.1$.\\
\indent$\bullet{}$~\textbf{Convective Mixing:}~We use the time-dependent convection scheme of Ref.~\cite{1986A&A...160..116K} with mixing length $\alpha_{\rm MLT}=1.8$.~This scheme includes the effects of turbulent convection and reduces to the Cox \& Giuli scheme \cite{1968pss..book.....C} at late times.\\
\indent$\bullet{}$~\textbf{Nuclear Reaction Network}:~We use the MESA {\sc pp\_cno\_extras\_o18\_ne22.net} network.\\
\indent$\bullet{}$~\textbf{Opacity}:~We use the Asplund et al.~(2009) type-II opacity tables \cite{2009ARA&A..47..481A}.\\
\indent$\bullet{}~$\textbf{Overshooting}:~We use the exponential scheme.\\

Our implementation of DM annihilation is as follows.~Our code takes as input the DM mass, cross section, and ambient density following the assumptions in the previous section.~This specifies the DM spatial profile and its normalization.~We write Equation~\eqref{eq:energyinjection} (having substituted Equation \eqref{eq:ndm_interpol}) as:
\begin{equation}\label{eq:eps(r)redef}
\varepsilon(r)=\varepsilon_0\frac{f(r)^2}{\rho(r)}
\end{equation}
with 
\begin{align}
    \varepsilon_0&=\langle\sigma v\rangle \mdm n(0)^2;\quad \textrm{and}\\
    f(r)&= \frac{n_{\rm DM}(r)}{n_{\rm DM}(0)}.
\end{align}
With these definitions, Equation \eqref{eq:lumanneq} can be rearranged to find $\varepsilon_0$
\begin{equation}\label{eq:epsilon0def}
    \varepsilon_0=\frac{R^2\Phi\mdm}{4\chi};\quad\chi=\int_0^{R}r^2f(r)^2\mathrm{d}r,
\end{equation}
where we took $f_\nu=1$ and $\Gamma_{\rm cap}=\pi R^2\Phi$;~$\Phi$ is given in Equation \eqref{eq:DMflux}.~At each time step, we calculate $\chi$ by integrating over the stellar profile and use Equation \eqref{eq:epsilon0def} to calculate $\varepsilon_0$.~We then use this to calculate the injection in each cell using Equation~\eqref{eq:eps(r)redef}.

 \section{Blue Loop Evolution Under Dark Matter}
 \label{sec:blue_loops_DM}

\begin{figure}
    \centering
    \includegraphics[width=0.48\textwidth]{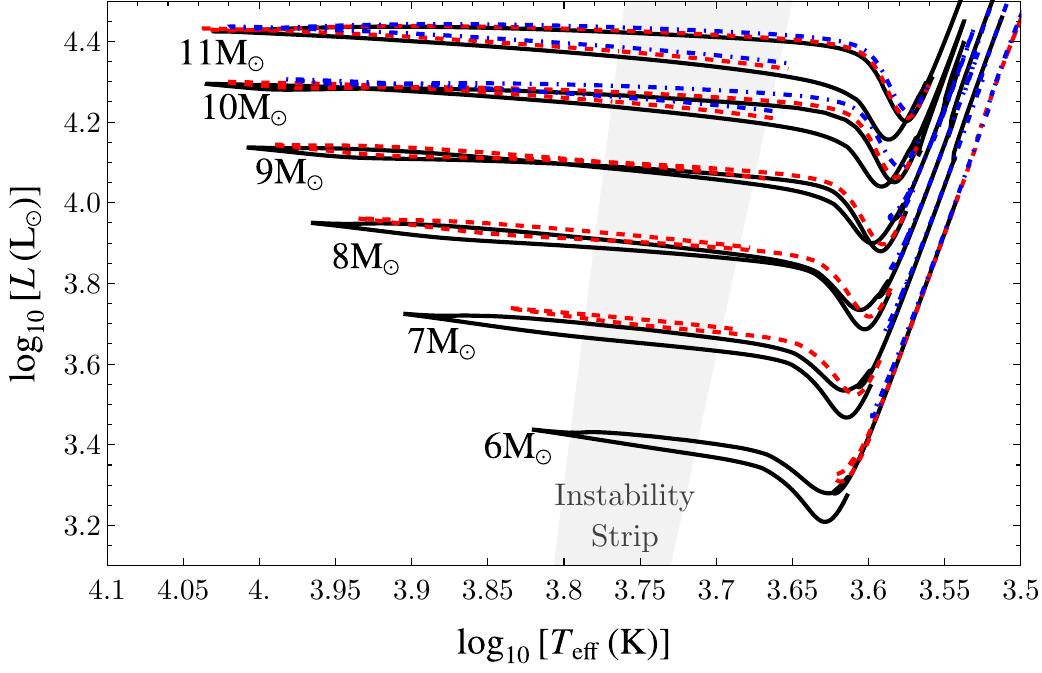}
    \caption{Evolution of $6$--$11~\msun$ stars for the SM (black, solid) and under DM annihilation with $m_{\rm DM}=1$ GeV, $v_{\rm DM}= 50$ km/s, and $\vtwo{f_{\rm cap}}\,\rho_{\rm DM}=10^5$ GeV/cm$^3$ (red, dashed) \vtwo{as well as $f_{\rm cap}\,\rho_{\rm DM}=2\times 10^5$ GeV/cm$^3$ (blue, dot-dashed)}.~The gray region shows the instability strip.
    }
    \label{fig:DM_loops}
\end{figure}

We simulated a grid of stars with masses from $5$--$11~\msun$.~The initial metallicity and helium abundance were set to $Z=0.02$ and $Y=0.29$, typical of galactic center stars \cite{2007ApJ...669.1011C,2009ApJ...691.1816N,2022MNRAS.513.5920F,2022yCat..75135920F}.~The DM mass and circular velocity were set to \mbox{$m_{\rm DM}=1$ GeV} and \mbox{$v_{\rm DM}= 50$ km/s}.\vtwo{~We varied the captured DM density around the fiducial value \mbox{$\vtwo{f_{\rm cap}}\rho_{\rm DM}=10^5$ GeV/cm$^3$}.~}~The resulting evolution in the Hertzsprung-Russell (HR) diagram is shown in Figure~\ref{fig:DM_loops} \vtwo{for two different capture densities}, with a zoom-in on the 6–7~$\msun$ range in Figure~\ref{fig:hold_my_beer}\vtwo{, where we focus on $\vtwo{f_{\rm cap}}\rho_{\rm DM}=10^5$ GeV/cm$^3$}.

\begin{figure}
    \centering
    \includegraphics[width=0.4\textwidth]{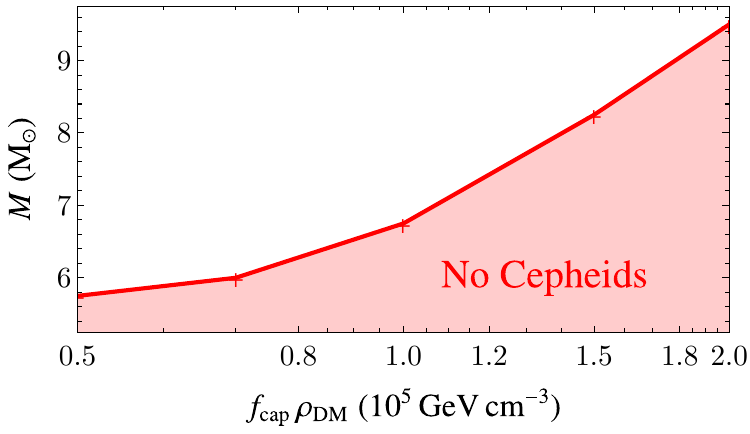}
    \caption{The dependence of Cepheid formation on the capture density $f_{\rm cap} \rho_{\rm DM}$.~ 
    }
    \label{fig:densitydep}
\end{figure}

At fixed stellar mass, the blue loops of stars evolving with DM annihilation are suppressed relative to the Standard Model prediction:~the stars remain cooler.~The suppression is strongest at lower stellar masses, as DM heating competes with nuclear burning, which dominates in massive objects.~For example, a $7~\msun$ star still executes a loop and crosses the instability strip, though at lower effective temperature, while a $6~\msun$ star fails to develop a loop and does not reach the Cepheid phase.

Exploring other DM parameters confirms that this behavior is generic.~Increasing $\vtwo{f_{\rm cap}}\rho_{\rm DM}$ strengthens the effect, suppressing loops even in higher-mass stars.\vtwo{This can be seen in Fig.~\ref{fig:DM_loops}, where the higher DM density of $ f_{\rm cap } \rho_{\rm DM} = 2 \times 10^5 \, \rm GeV \, cm^{-3}$ implies the absence of instability strip crossing by stars up to $10 \rm \, M_\odot$, whereas for $ f_{\rm cap } \rho_{\rm DM} = 1 \times 10^5 \, \rm GeV \, cm^{-3}$, the strip is crossed by $7 \, \rm M_\odot$ stars.~We show the dependence of this lightest mass on the capture density in Fig.~\ref{fig:densitydep}.}~Decreasing $v_{\rm DM}$ enhances capture through gravitational focusing (see Eq.~\eqref{eq:DMflux}), producing the same trend.\vtwo{~In the limit $v_{\rm esc} \gg v_{\rm DM}$ relevant to Cepheid stars in the galactic center, changes in $v_{\rm DM}$ are approximately degenerate with changes in $ (f_{\rm cap } \rho_{\rm DM} )^{-1}$.}~Changing $m_{\rm DM}$ does not qualitatively alter the results since the amount of injected energy is independent of $m_{\rm DM}$.

\begin{figure}[t!]
    \centering
    \includegraphics[width=0.5\textwidth,height=3.8cm]{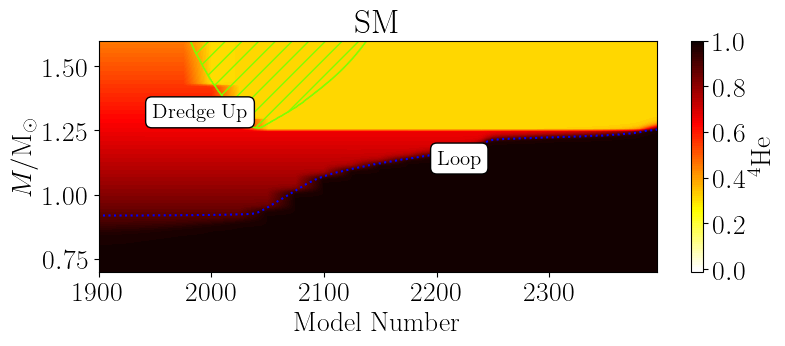}
    \includegraphics[width=0.5\textwidth,height=3.8cm]{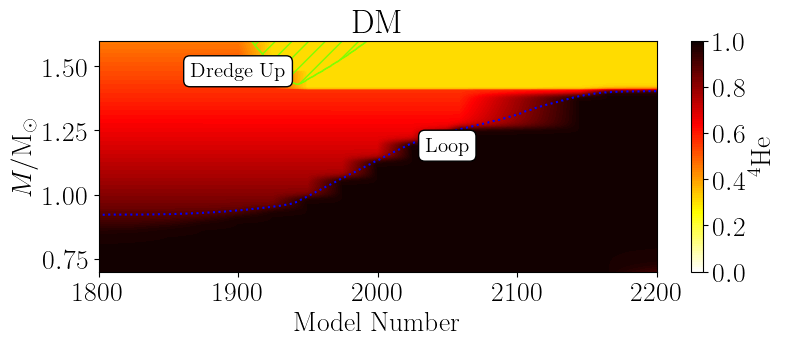}

    \caption{Kippenhahn diagrams showing the $^4$He mass fraction for $6~\msun$ stars during the dredge up and the start of the blue loop.~The x-axis proceeds monotonically in time, and the y-axis shows the enclosed stellar mass within a shell.~Both a SM star (upper) and a star evolving under DM energy injection (lower) are shown.~The DM plot uses $m_{\rm DM}=1$ GeV, $\rho_{\rm DM}=10^5$ GeV/cm$^3$, and $v_{\rm DM}=50$ km/s.~The blue dashed lines indicate the location of the helium core.}
    \label{fig:kipps_effect}
\end{figure}

The blue loop suppression in Figures~\ref{fig:hold_my_beer} and \ref{fig:DM_loops} can be explained by examining the Kippenhahn diagrams in Figure~\ref{fig:kipps_effect}, which show the mass fraction of helium-4 
for stars with and without DM annihilation.~The horizontal axis shows the model number, which increases sequentially as the stellar evolution calculation proceeds.~The mapping between model number and physical time is non-linear, since the timestep size varies adaptively depending on the evolutionary phase and numerical stability requirements.~The vertical axis shows the mass coordinate, i.e., the radial coordinate expressed as the mass enclosed within a given shell of the star, in units of solar mass.~The plots are zoomed in to highlight the inner regions where mixing events and structural changes occur, rather than displaying the full stellar mass.
~During the blue loop phase, a hydrogen burning shell surrounding the core provides both the hydrostatic support for the star and its luminosity.~The looping is the result of the excess helium above the shell \cite{2015MNRAS.447.2951W}, which makes the star larger and redder than it would otherwise be.~As the shell burns outwards, the excess is reduced causing the envelope to become hotter, driving the envelope to contract and the star to shift blueward in the HR diagram.~This terminates when core helium burning ceases and the core contracts causing the envelope to inflate and cool so that the star returns to the Hyashi line --- the redward motion.~Evidently, the DM object has excess helium extending to larger mass coordinates compared with the SM object so the shell must burn further out before the envelope can be heated.~This delay reduces the time available for blueward motion before core helium exhaustion.~Consequently, under DM annihilation the star performs a less extended loop.

\begin{figure*}
    \centering
    \includegraphics[height=3.85cm]{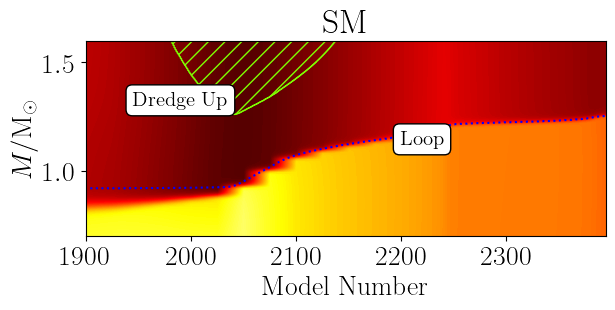}
    \includegraphics[height=3.85cm]{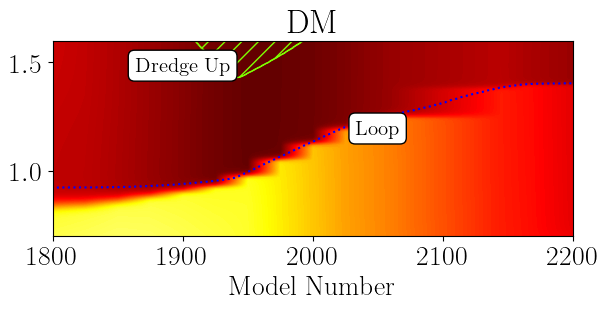}
    \includegraphics[width=1.5cm,height=3.25cm]{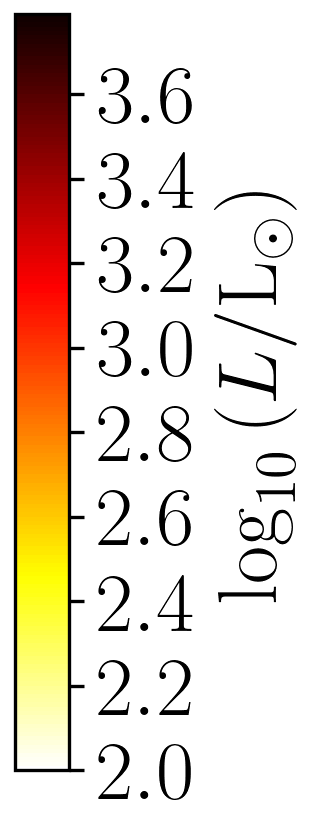}
    \\
    \includegraphics[height=3.5cm]{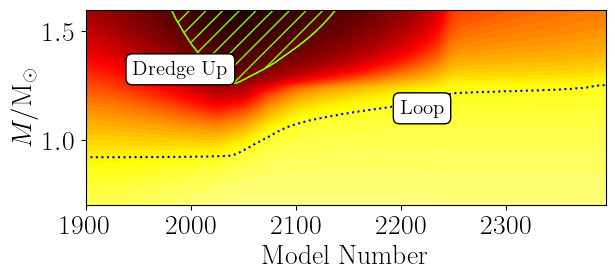}
    \includegraphics[height=3.5cm]{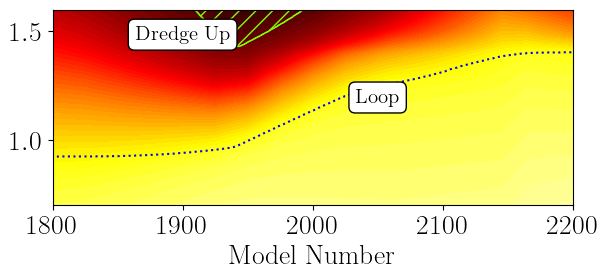}
    \includegraphics[width=1.5cm,height=3.25cm]{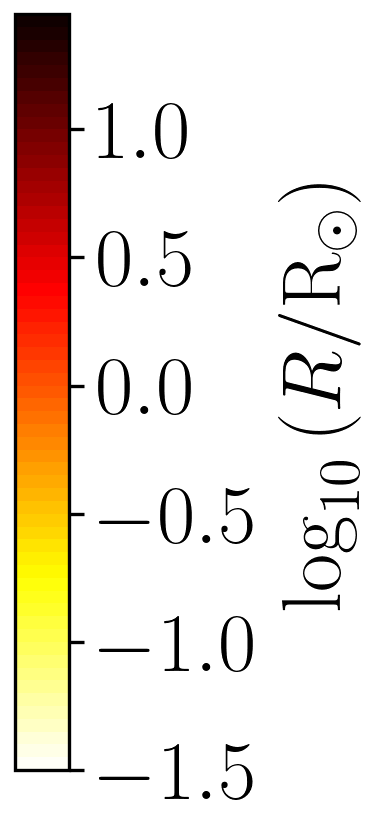}
    \\
    \includegraphics[height=3.5cm]{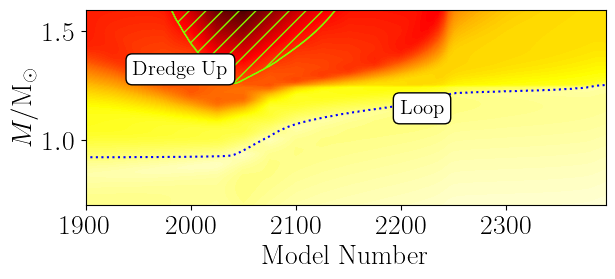}
    \includegraphics[height=3.5cm]{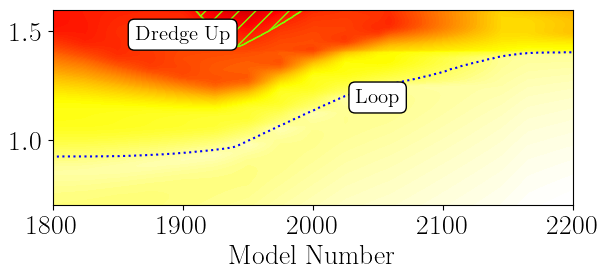}
    \includegraphics[width=1.5cm,height=3.25cm]{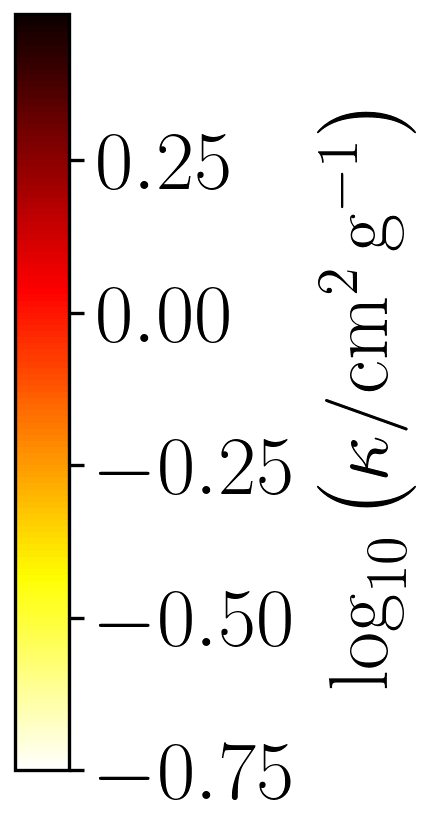}
    \caption{Kippenhahn diagrams for $6~\msun$ models around first dredge-up.~Left:~SM; right:~with DM heating ($m_{\rm DM}=1$\,GeV, $\vtwo{f_{\rm cap}}\rho_{\rm DM}=10^5\,{\rm GeV\,cm^{-3}}$, $v_{\rm DM}=50$\,km\,s$^{-1}$).~Blue dashed lines mark the He-core boundary.}
    \label{fig:kipps_cause}
\end{figure*}

The extended helium excess is due to differences during the first dredge up (point D in Fig.~\ref{fig:cepheid_evolution}).~After the main-sequence, the star burns hydrogen in a thin shell around the core.~This hydrogen shell burning drives the envelope to expand and cool, increasing the opacity and making radiative transport less efficient.~As a result, convection becomes the dominant energy transport method.~The convection zone extends inward, reaching the core and pulling up helium and other CNO products from deeper layers to the surface.~In the DM model, the convection zone does not penetrate as deep, which means less helium is pulled toward the surface.~This results in a larger excess of helium above the shell during the blue loop.

The Kippenhahn diagrams in Fig.~\ref{fig:kipps_cause} illustrate how DM heating leads to a shallower convective penetration.~DM annihilation provides an additional source of pressure support, so the hydrogen-burning shell contributes less to maintaining hydrostatic equilibrium.~This is evident in the upper panels of Fig.~\ref{fig:kipps_cause}, where the shell luminosity is reduced in the DM model.~With less energy injected into the overlying layers, the envelope expands less and the temperature at its base remains higher.~The middle row of Fig.~\ref{fig:kipps_cause} shows this explicitly through the radial coordinate.~The hotter base temperature lowers the opacity (since $\kappa \propto \rho/T^{3.5}$), making radiative transport more efficient and limiting the depth of the convective zone, as seen in the bottom panels of Fig.~\ref{fig:kipps_cause}.

\section{Discussion and Conclusions}
\label{sec:discussion}

In this work, we investigated the impact of dark matter annihilation on the evolution of Cepheid variable stars.~We found that for captured DM densities above $10^5~{\rm GeV\,cm^{-3}}\vtwo{/f_{\rm cap}}$, the blue loop phase is suppressed, preventing the formation of short-period ($1$--$6$ day) Cepheids.~Such densities are possible in the inner parsec of the Milky Way, making this region a natural testing ground for our prediction. \vtwo{Slightly larger values for the captured DM densities (e.g., $2 \times 10^5~{\rm GeV\,cm^{-3}}\vtwo{/f_{\rm cap}}$) significantly increase the mass range of Cepheids that are affected.}

Our results above imply that low-mass Cepheids, those with masses $M\le6~\msun$, cannot form within the inner parsecs of the galactic center \vtwo{within our default model}.~These objects have short-periods spanning $1$--$6$ days, so their absence would be consistent with expectations from WIMP DM heating.

We note that no Cepheids have yet been discovered in regions close enough to the galactic center to probe or constrain dark matter effects.~There are several factors that explain the lack of existing Cepheid observations.~Astrophysically, the initial mass function in the GC is top-heavy, reducing the fraction of 3--12~M$_\odot$ stars compared to the larger O-star population~\cite{2010ApJ...708..834B}.~In addition, the confirmed star-formation history is sparse:~the most secure event occurred $\sim$6 Myr ago, too recent to produce Cepheids, though additional less well-established episodes of star-formation have been suggested~\cite{2017MNRAS.469.2263B}.~Observationally, crowding, extinction, and diffuse backgrounds hinder their detection.~However, this latter issue is likely to be solved by the extreme sensitivity of near-future near-infrared (NIR) telescopes that are capable of detecting these systems.

Assuming Cepheids did form, a top-heavy IMF with slope $\alpha = 0.45$--$1.7$ (indicated by GC observations \cite{2010ApJ...708..834B,2013ApJ...764..155L}) predicts that $45$--$64\%$ of the expected population ($3$--$11~\msun$) would be missing due to DM annihilation, \vtwo{in scenarios with our fiducial DM density of 10$^5$~GeV~cm$^{-3}$.}~Thus, even under favorable astrophysical conditions, the selective absence of short-period Cepheids constitutes a potential signature of WIMP dark matter heating. \vtwo{In scenarios with even marginally higher DM densities, the blue loop can be suppressed across the entire Cepheid mass range, leading to the non-detection of Cepheids in the galactic center. It is worth noting that these effects may be combined due to the sharply rising DM densities in the central region of the Milky Way, producing a radially dependent maximum Cepheid mass that would be indicative of DM contributions as it would be difficult to fit via only astrophysical mechanisms.}

\vtwo{Should the lack of observed Cepheids at the galactic center persist into the JWST/NIRCam and ELT/MICADO era, the degeneracies mentioned in Sec.~\ref{sec:Cepheid_intro} motivate a systematic exploration of stellar physics uncertainties in conjunction with DM heating.~Such a multidimensional study is computationally intensive and beyond the scope of this work, whose aim is to establish the existence and magnitude of the DM-induced suppression mechanism.
}

\section*{Software}
MESA version~23.05.01, MESASDK version 23.7.3, Asteria version 1, Mathematica version 12, Jupyter Notebook 6.4.12, Python version 3.8.5.

\section*{Acknowledgements}
We are grateful for discussions with Dan Hey and David Rubin.~DC is supported by the STFC under Grant No.~ST/T001011/1, and is grateful to the Mainz Institute for Theoretical Physics (MITP) of the Cluster of Excellence PRISMA+ (Project ID 390831469) for its hospitality and its partial support during the completion of this work.~TL acknowledges support by the Swedish Research Council under contract 2022-04283.~This material is based upon work supported by the National Science Foundation under Grant No.~2207880.~Our simulations were run on the University of Hawai\okina i's high-performance supercomputer KOA.~The technical support and advanced computing resources from University of Hawai\okina i Information Technology Services – Cyberinfrastructure, funded in part by the National Science Foundation MRI award \#1920304, are gratefully acknowledged.~

\bibliography{refs}

\end{document}